\documentclass[twocolumn,10pt]{revtex4-1} 
\usepackage{amsmath,amssymb,graphicx}
\usepackage{color}
\usepackage{soul}
\usepackage{ulem} 

 \graphicspath{ {./} }
\begin{document}

\title{Two-dimensional materials and the coherent control of nonlinear optical interactions}

\author{Shraddha M. Rao$^1$, Ashley Lyons$^1$, Thomas Roger$^1$, Matteo Clerici$^1$, Nikolay I. Zheludev$^2$, and Daniele Faccio$^1$
}
\email{d.faccio@hw.ac.uk}
\address{$^1$School of Engineering and Physical Sciences, SUPA, Heriot-Watt University, Edinburgh EH14 4AS, UK\\
$^2$Optoelectronics Research Centre, University of Southampton, Southampton SO17 1BJ, UK
}

\date{\today}

\begin{abstract}
Deeply sub-wavelength two-dimensional films may exhibit extraordinarily strong nonlinear effects. Here we show that 2D films exhibit the remarkable  property of a phase-controllable nonlinearity, i.e., the amplitude of the nonlinear polarisation wave in the medium can be controlled via the pump beam phase and determines whether a probe beam will ``feel'' or not the nonlinearity. This is in stark contrast to bulk nonlinearites where propagation in the medium averages out any such phase dependence. We perform a series of experiments in graphene that highlight some of the consequences of the optical nonlinearity phase-dependence, {such as} the  coherent control of nonlinearly diffracted beams, single-pump-beam  {induced} phase-conjugation and the demonstration of a nonlinear mirror characterised by negative reflection. The observed phase sensitivity is not specific to graphene but rather is solely a result of the  dimensionality and is therefore expected in all 2D materials.

\end{abstract}
  

\maketitle
Two-dimensional optical media with deeply sub-wavelength or mono-atomic thickness such as photonic metamaterials, plasmonic, heterostructure layered materials, and structured or few-layer graphene provide the advantage of being able to  control (enhance, suppress or modulate) optical interactions. Their functionality in the linear regime is well established in terms of cloaking \cite{schruig}, ultrafast modulators \cite{zheludev2} and optical magnetism. Their nonlinear functionalities too are being explored, such as four wave mixing (FWM) and in particular, phase-conjugation and optical negative refraction, i.e. a phase conjugated field that is transmitted, rather than reflected as in the standard situation and, moreover, at negative angles \cite{pendry}. Such negatively refracted beams have been used to perform perfect imaging \cite{alu}.\\
Graphene is a promising optical material with unique properties related mostly to the material linear disperison in vicinity of the  Dirac points, e.g.,  high yet constant absorption over a huge bandwidth \cite{geim}, electrically controllable optical properties \cite{smith} and high electron mobility. It is widely studied as a saturable optical absorber \cite{bao} but only more recently has started to attract attention for its third order nonlinear properties \cite{novotny}. Of particular relevance is the recent demonstration of an 8-order of magnitude enhancement of the $\chi^{(3)}$ with respect to typical dielectric materials \cite{hendry}.\\
However, similar to bulk 3D nonlinear interactions, the 2D nonlinearity has been treated as a ``phase independent'' effect that is not affected by the relative phases between the interacting input beams. For example, four-wave mixing will ensue regardless of the relative phases between  the input pump beam (with phase $\phi_p$) and input probe or signal beam (with phase $\phi_s$)  so that an ``idler'' wave will always be generated and will acquire the relative phase difference $\phi_i=2\phi_p-\phi_s$. Even in the case in which a phase-dependence is known to arise, i.e. when all four beams (i.e. two pumps, a  signal and an idler) are simultaneously incident on the nonlinear medium, a nonlinear polarisation is still generated and will ensure photon-photon interactions, i.e. energy flow between the beams where the relative pump/signal/idler phases only determine the direction in which the energy flows. In this sense, the specific nonlinear process (characterized by the direction of energy flow) is phase-dependent but the (presence of the) nonlinearity itself is not. \\
Conversely,  we may also envisage a ``phase-dependent'' nonlinearity: in this case, a nonlinear polarisation wave in the medium is physically excited (or not excited) only for certain relative phases of the input beams. For example,  if the pump beams have the same phase then nonlinear interactions and four-wave mixing will occur and if they have opposite phases, the nonlinear polarisation is not excited. In the latter case, there will be no possibility for photon-photon interactions and the probe beam will therefore propagate as if in a purely linear medium. Access to such a phase-control of the  nonlinearity  would extend the applications and control currently available in nonlinear optics: some ideas in this sense are demonstrated in the following.\\
\begin{figure}[!t]
\begin{center}
\includegraphics[width=9cm]{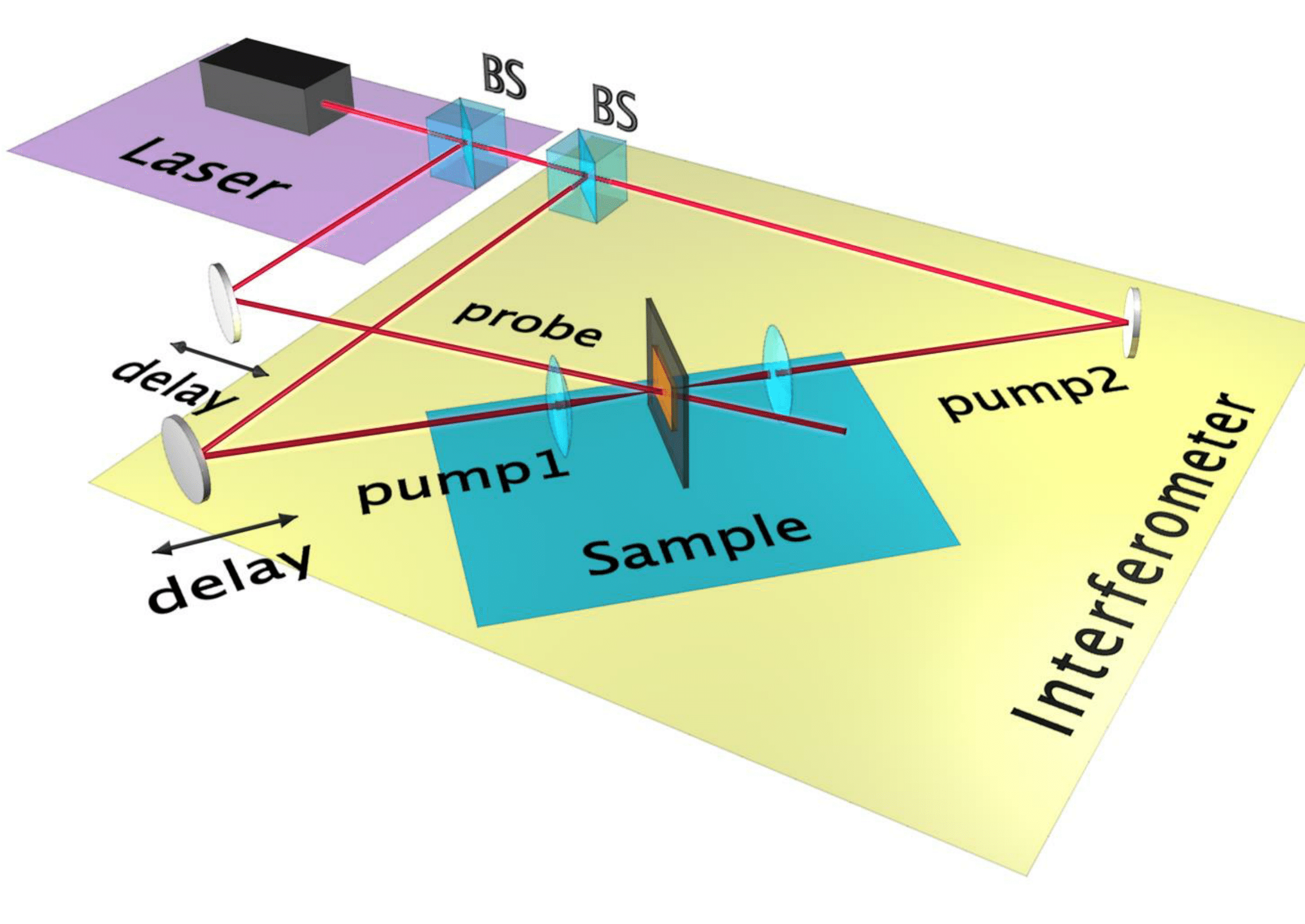}
\caption{A schematic representation of the experimental layout: the counterpropagating pump beams ({pump 1} and {pump 2}) and a probe beam, generated from a Ti:Sapphire laser (100 fs pulses centred at 780 nm), overlap on a 30-layer graphene sample. The two pump beams counterpropagate in a Sagnac interferometer loop and the arrangement allows to finely tune the relative phase of pump 1 with respect to pump 2 with a piezoelectric stage-controlled mirror. A piezoelectric actuator is placed also on the probe arm.}
\label{layout}
\end{center}
\end{figure}
We show how two-dimensional films naturally support coherent phase control of the nonlinearity with properties that are not found in bulk 3D materials. This in turn enables the coherent control (i.e. on/off modulation through the phase of the signal or pump beam) of known processes such as optical phase-conjugation but can also can give rise to novel nonlinear wave-mixing geometries. In particular, optical phase conjugation and diffraction at negative angles (a.k.a {\emph{optical} negative refraction) are shown to occur even in the presence of a single pump beam (as opposed to two counter-propagating beams). These are just some examples of the opportunities provided by the phase-control of the  nonlinearity.  \\

Experiments are carried out using a 100 fs pulse duration, 100 Hz repetition rate, $\lambda=780$ nm wavelength Ti:sapphire laser. The experimental setup is composed of a Sagnac-interferometer that enables coherent interaction between the two counterpropagating pump beams at the sample, as shown in Fig.~\ref{layout}. The typical energy of each pump beam on the graphene sample is of the order of 1 $\mu$J and the beams are focused down to diameters of the order 50 $\mu$m, i.e. the laser intensities on the graphene film are of the order of 100 GW/cm$^2$. The nonlinear sample is a multilayer (30 layer) graphene, of thickness $\sim9$ nm, sandwiched between two fused-silica layers.  A  probe beam with the same wavelength and spatial width is overlapped onto the pump beams  at an angle of 1.8 deg. (see Fig.~\ref{fig2}).  The temporal overlap is optimised with a tunable optical delay line on the probe beam-path.  The phase of one of the pump beams on the sample is then finely controlled by a piezoelectric stage mounted on the last mirror that directs the pump beam onto the sample. This layout is essentially the same as that used to demonstrate the coherent modulation of the linear properties of deeply subwavelegnth films, e.g. absorption as demonstrated by Zhang et al. \cite{zheludev} using metamaterials and more recently also using graphene \cite{rao}.
\\

\begin{figure}[!t]
\begin{center}
\includegraphics[width=8cm]{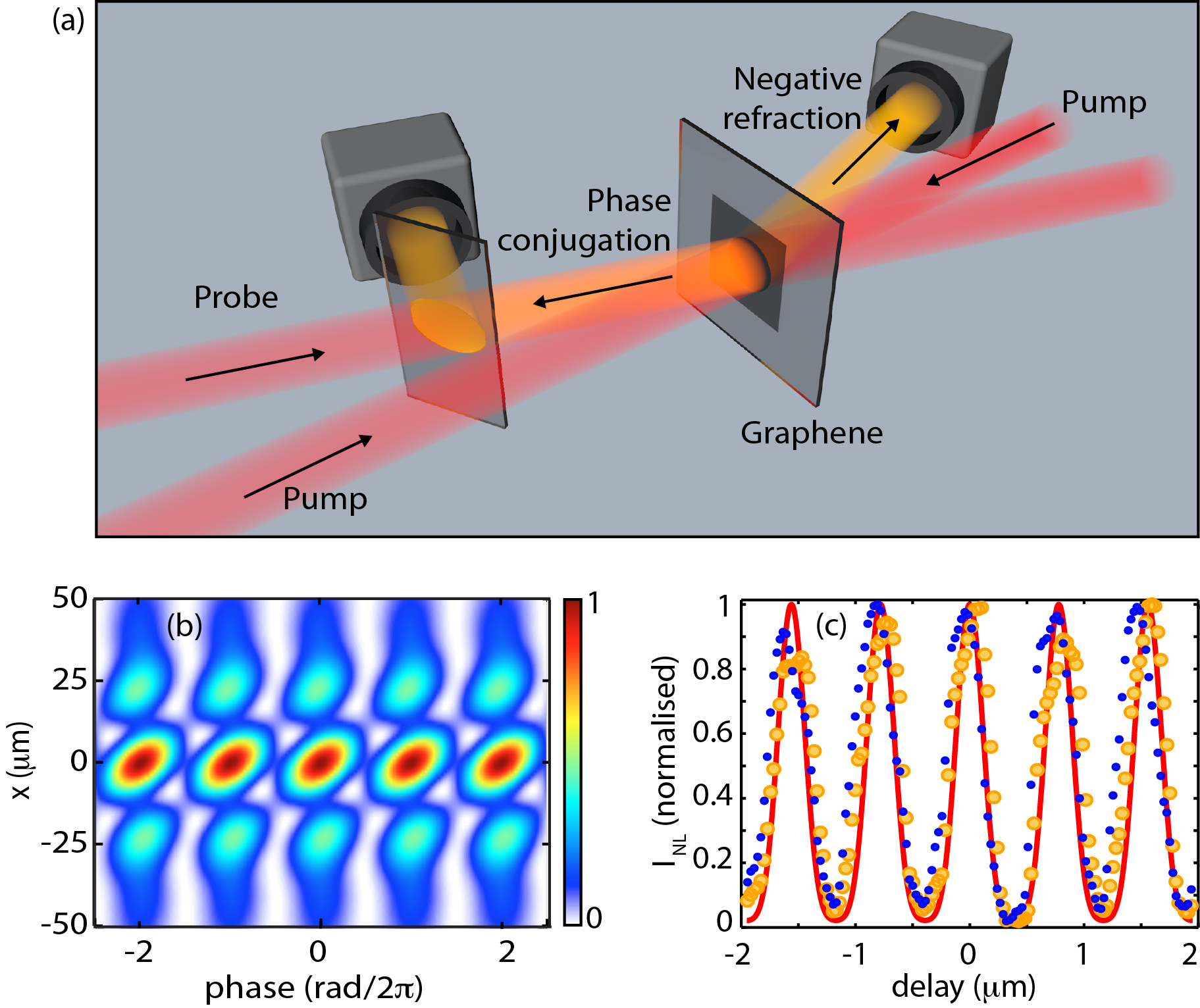}
\caption{Generation and modulation of nonlinear signals in the three-beam geometry. (a) A schematic representation of the generation of the phase conjugated and the negatively diffracted signals (shown in orange) when the two counterpropagating pump beams and the probe beam are in phase.  (b) shows the full spatial distribution on the graphene film of the {time-averaged} nonlinear polarisation  $P_{nl}^2$, as a function of the relative {phase} delay between pump 1 and pump 2. (c) shows the phase-conjugated (blue dots) and negatively refracted (yellow circles) beam  amplitudes as a function of the  {relative delay between pump 1 and pump 2}. The red curve is the theoretical calculation of {time-averaged} $P_{nl}^2$ as described in the text.}
\label{fig2}
\end{center}
\end{figure}
\emph{Three-beam interaction}\\
 The first experimental configuration involves measurements where the graphene sample is illuminated with both counterpropagating pump beams and the probe beam (all having the same frequency $\omega$), as shown in Fig.~\ref{fig2}(a). Although all three beams have the same input energy ({5 $\mu$J}) and intensity ({100 GW/cm$^2$}), for descriptive purposes we indicate as ``pump'', any beams that are orthogonal to the graphene film and as ``probe'', any beams that are incident at an angle with respect to the surface normal. The counterpropagating pump beams therefore form a standing wave on the  graphene film. Degenerate Four Wave Mixing (DFWM) between the three beams creates a phase conjugated  signal that propagates reflected exactly along the input probe  and is separated from the latter by a partially reflecting beam-splitter.  In addition to this  signal, we also observe a beam that is transmitted through the sample and, with respect to the input probe beam, is negatively refracted. Energy conservation dictates that both of these beams also have the same frequency $\omega$ as the input beams. The origin of the negatively refracted beam, reported in graphene for the first time in Ref.~\cite{novotny}, lies in the fact that for such thin films all propagation effects and in particular all phase-matching constraints involving the wave-vector components along z (the propagation direction) that dominate the nonlinear beam evolution in bulk media, here become irrelevant. Hence, the scattering of an input probe beam into an output beam arising from the nonlinear polarisation wave depends only on the transverse components of the wave-vector components, $k_\perp^{in}$ and $k_\perp^{out}$, respectively. If the two counterpropagating pump beams are at normal incidence on the film, then momentum conservation implies $k_\perp^{out}=-k_\perp^{in}$, i.e. the output beam appears as if it is propagating in a medium with an \emph{effective} refractive index $n=-1$. {We note that the four-wave mixing process responsible for the negatively diffracted beam  has been referred to in the past as \emph{forward degenerate four-wave mixing} or \emph{forward phase conjugation} (see e.g. \cite{FWMreview}) and has been for instance employed for the characterization of resonant nonlinarities in multiple quantum wells \cite{MQW}. However, in recent literature this same effect is often referred to as ``{\emph{optical} negative refraction'' (e.g. \cite{palomba,novotny}) in relation to the fact that the medium behind the nonlinear 2D surface behaves to all effects as if it where a negative index medium \cite{pendry}.\\
A marked difference with respect to previous measurements is the phase-dependence of the nonlinearity, which becomes evident  by controlling the relative phase of the two pump beams. The amplitude of the nonlinear polarisation wave that is responsible for emission of the phase conjugated and negatively refracted signals is given by: {$P_{nl}\propto \chi^{(3)}(\mathcal{E}_\textrm{pump 1}+\mathcal{E}_\textrm{pump 2}+\mathcal{E}_\textrm{probe})^3$, where $\mathcal{E}$ is the real electric field. It is hence clear that, at a given propagation coordinate, selected by the sub-wavelength film,} for a fixed phase of two of the beams {the nonlinear polarisation} depends on the phase of the third one.  {In Fig.}~\ref{fig2}(b) {we report the time-averaged}  $P_{nl}^2$ {for a chosen} film position that shows the phase-dependence in the form of clear oscillations with respect to the {relative} phase {between} $E_\textrm{pump 1}$ {and $E_\textrm{pump 2}$, } { which determines} whether DFWM will occur (with the generation of a phase conjugated and negatively refracted beam) or not. \\
Another way of viewing this is to write the  polarisation {component that is responsible for the generation of the optical negative refraction and the phase conjugation} as { $P=\varepsilon_0[\chi^{(1)}+2\chi^{(3)}(\mathcal{E}_\textrm{pump 1}+\mathcal{E}_\textrm{pump 2})^2]\mathcal{E}_\textrm{probe}$}, where we have isolated the polarisation term that is linear in $\mathcal{E}_\textrm{probe}$ and neglected all third harmonic and nonlinear mixing terms that involve only the two other (pump) beams. This formula shows that the two pump beams create, at a fixed propagation coordinate, an effective film susceptibility $\chi_\textrm{eff}=\chi^{(1)}+2\chi^{(3)}(\mathcal{E}_\textrm{pump 1}+\mathcal{E}_\textrm{pump 2})^2$ that oscillates at twice the pump frequency and whose amplitude is determined by the relative pump beam phases. 
This viewpoint highlights the phase dependence of the material susceptibility, which in turn determines whether the incoming probe beam encounters a ``linear'' or a ``nonlinear'' medium. In Fig.~\ref{fig2}(b) we plot the {time-averaged} $P_{nl}^2$ calculated at the film position, which can be clearly seen to oscillate as result of the  $\chi_\textrm{eff}$ phase-sensitivity. \\ 
In the experiments the phase of pump 1 (Fig.~\ref{layout}) is precisely varied with respect to that of pump 2 using a piezoelectric-controlled mirror: the amplitude of the phase conjugated and negatively refracted beams are recorded on  photodetectors as a function of the piezoelectric-mirror position. A periodic  modulation in the amplitudes of the nonlinear signals is observed with a close to 100\% modulation amplitude as shown in Fig.~\ref{fig2}(c): the blue dots represent the signal amplitudes of the phase conjugate and the optical negative refraction signals, respectively. The red solid line is the normalised plot of {the time-averaged} $P_{nl}^2$ without any free parameters, other than an overall arbitrary phase that has been adjusted to overlap with the data. \\

\begin{figure}[!]
\begin{center}
\includegraphics[width=8cm]{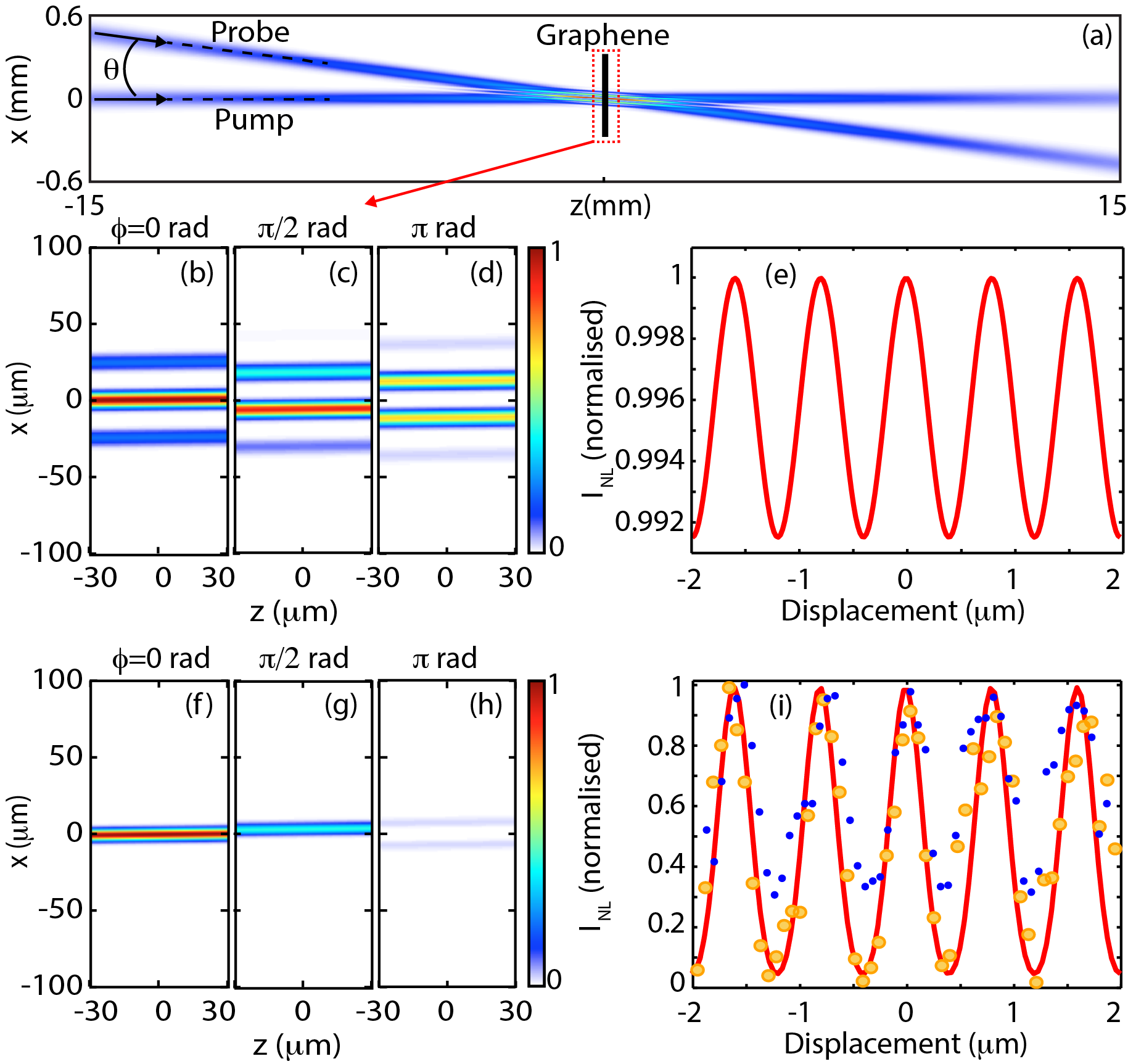}
\caption{Two beam interaction: (a) shows the pump and probe beam layout with respect to the graphene film. (b), (c) and (d) are zoomed in images of the calculated transverse (x) and longitudinal (z) distribution of the {time-averaged} nonlinear polarisation $P_{nl}^2$ for different relative phases, $\phi$, of $E_\textrm{pump}$ and $E_\textrm{probe}$ as indicated in the figure. The two beams have the same diameter of 80 $\mu$m and generate several spatial interference fringes. The calculated intensity of the negatively refracted beam is shown in (e) as a function of relative phase delay $\phi$: only a very weak modulation is observed. (f), (g) and (h) show the same as in (b)-(d) but with smaller beam diameters of 20 $\mu$m. (i) shows the experimental data for the measured negatively refracted (yellow circles) and phase conjugated (blue dots) beams as a function of piezo-mirror displacement. A strong $\sim100\%$ modulation is observed, in agreement with the theoretical calculation of time-averaged $P_{nl}^2$ (red curve).}
\label{fig3}
\end{center}
\end{figure}

\emph{Two-beam interaction}\\
In the second configuration of the experiment, we block one of the pump beams (pump 2) and study the interaction between pump 1 and the probe as function of the relative phase delay, see Fig.~\ref{fig3}(a).  {In this case, we still observe both the phase conjugation and  optical negative refraction, differently from the bulk case where  the longitudinal phase matching constraint suppresses these processes}.
Figures~\ref{fig3}(b-d) show the calculated  spatial distribution of the time-averaged nonlinear polarisation $P_{nl}^2$ at the graphene film when the spot size of the two beams is 80 $\mu$m for three different relative phase delays: as can be seen, changing the phase only leads to a lateral shift of the spatial interference pattern and under this condition, the DFWM nonlinear signals will be observed albeit with only a very weak modulation in their amplitudes  (see Fig.~\ref{fig3}(e) that shows the calculated {time-averaged} nonlinear polarisation, $P_{nl}^2$). However, as shown in Figs.~\ref{fig3}(f)-(h), if the spot size of the beams is reduced so that it is similar to the interference pattern fringe spacing, then changing the relative phase leads to  longitudinal \emph{and}  transverse modulation of the polarisation wave intensity, which in turn allows to coherently control the amplitude of the DFWM signals, similarly to the previous example. \\
\begin{figure}[!t]
\begin{center}
\includegraphics[width=8cm]{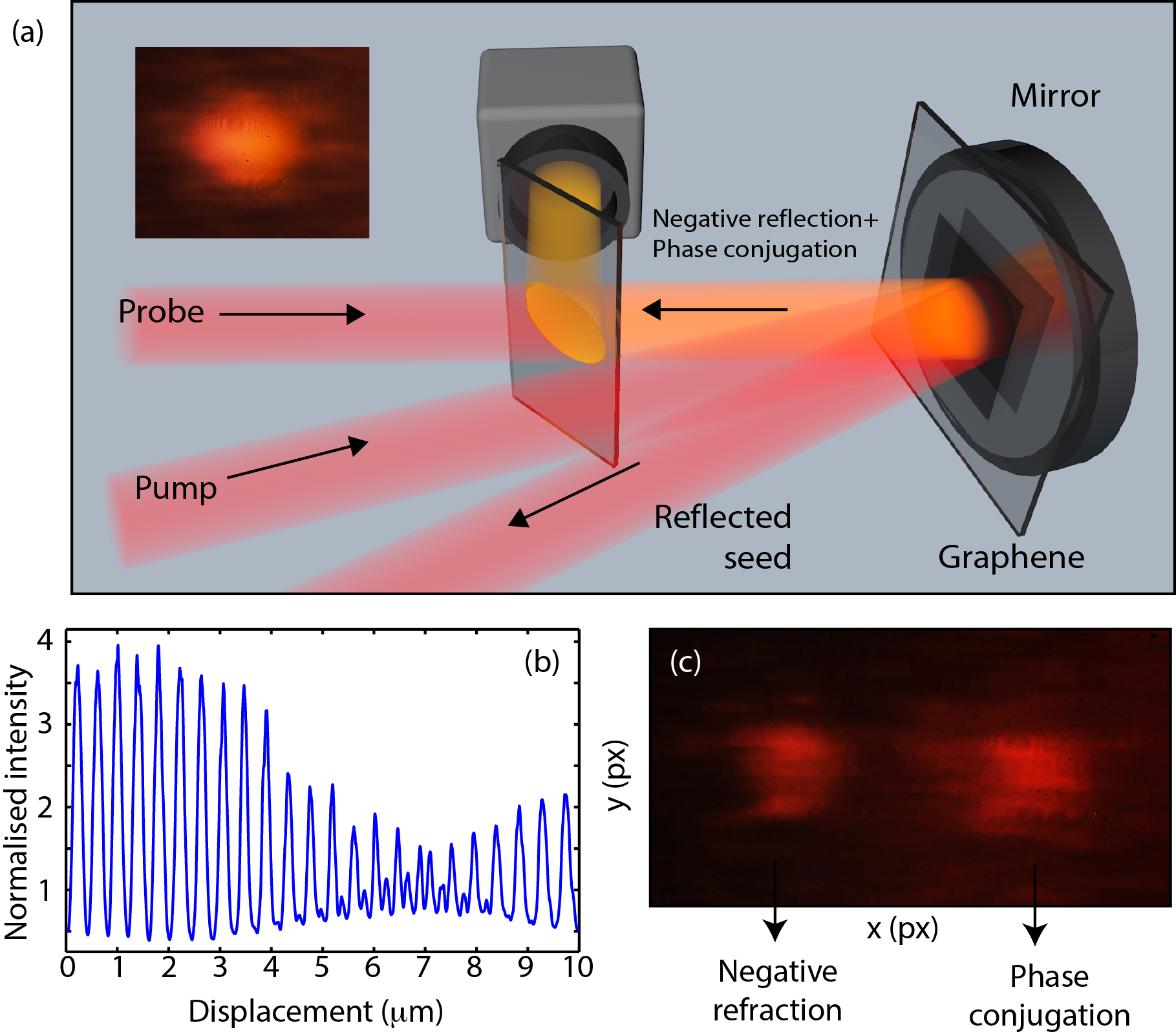}
\caption{Nonlinear coherent mirror: (a) Schematic representation of the experimental layout with a single pump beam at normal incidence to the graphene film and the probe beam at a small, 1.8 deg angle. A mirror is placed parallel to the film and its distance from the graphene film is controlled with a piezo-actuator. (b) Shows the recorded back-reflected signal (along the probe direction) as a function of mirror distance. (c) Shows the  spatial profile of the back-reflected signal recorded on a CCD camera with the mirror slightly detuned in angle. Two beams are observed: one is the phase-conjugated signal the other is the negatively refracted (and reflected back from the mirror) signal, as indicated in the figure. When the mirror is properly aligned parallel to the film, these two beams add coherently to form a single beam shown in the inset to (a).}
\label{mirror1}
\end{center}
\end{figure}
\begin{figure}[!t]
\begin{center}
\includegraphics[width=8cm]{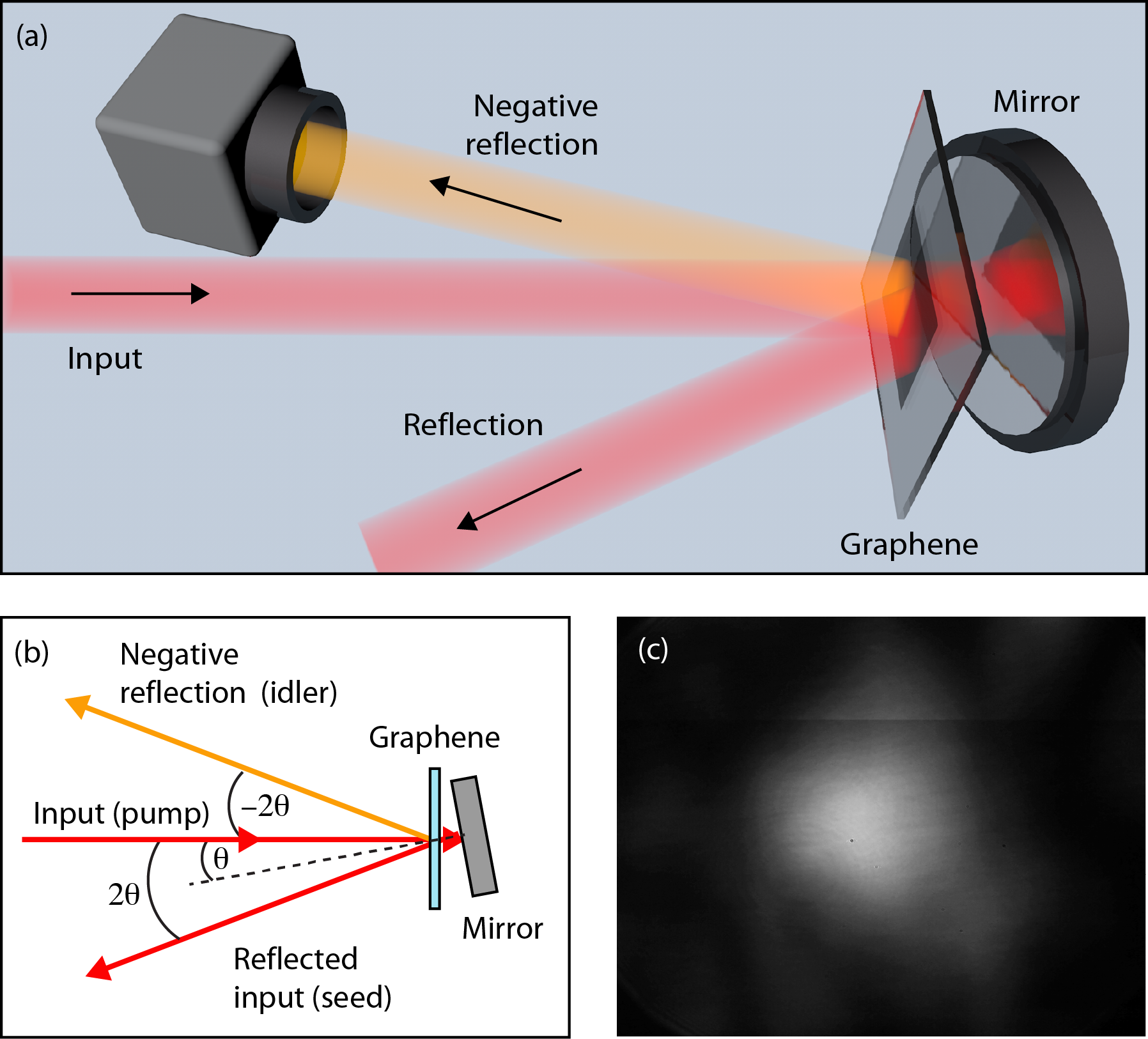}
\caption{Nonlinear mirror and negative reflection: (a) and (b) show a schematic representation of the experimental layout and wave-vector diagram with a laser beam incident normal to the graphene sample and with the mirror placed very closely to the film (at a distance shorter than the optical pulse length) but at a slight angle. Conservation of transverse momentum implies that the interacting beams generate a a reflected beam at an angle $2\theta$ (as expected from a standard mirror) but also a ``negative'' reflection at the opposite $-2\theta$ angle. (c) Shows an image of the negatively  reflected measured at angle $-2\theta$ on a CCD camera.}
\label{setup3}
\end{center}
\end{figure}
Experiments were carried out by reproducing the conditions of the calculations shown in Figs.~\ref{fig3}(f)-(h), i.e. by aligning the input beams with a relative 1.8 deg angle and by focusing them to a spot-size of 20 $\mu$m. Under this condition, we observe a periodic (with piezo delay) complete extinction of the negative refracted beam as shown in Fig.~\ref{fig3}(i), where the red dots represent the signal amplitude.  The phase conjugated beam exhibits a similar modulation (blue dots) albeit with a reduced visibility due to linear back-scattering of the pump signal from the sample surface, which in these experimental conditions proved difficult to completely eliminate. We note that phase conjugation is typically performed with two counterpropagating pump beams in bulk media but may be observed with a single beam when using a surface nonlinearity \cite{boyd}, as in our case. However, this is the first demonstration of  coherent control of the nonlinearity. We also note that, differently from the previous example where the modulation periodicity was given by the standing wave periodicity (i.e. $\lambda/2$), the amplitude of the nonlinear signals is now periodically modulated with a periodicity equal to the pump wavelength, i.e., $\lambda$ = 785 nm. Overall, this arrangement provides a simpler and equally effective method of generating an all-optical modulation of nonlinear signals in 2D films and may be extended also to the case of metamaterials, for example, and in general to perfecting imaging setups \cite{alu}.\\

\emph{Nonlinear coherent mirrors and negative reflection}\\
In this third example, we  perform the experiment in a reflection-configuration as shown in Fig.~\ref{mirror1}(a). The pump beam, focused using a lens with f = +50 cm, is at normal incidence to the graphene surface and the transmitted beam is reflected by a mirror placed closely behind and parallel to the film. The mirror distance from the film is reduced to be much smaller that the length of the optical pulse ($\sim30$ $\mu$m) and is fine-tuned using a piezoelectric stage. The reflected pump beam now acts a second pump and its relative phase with respect to the input beam is controlled by the mirror, thus allowing the same control over the resulting polarisation wave as in the first example above. Figure~\ref{mirror1}(b) shows the amplitude of the DFWM signal that is back-reflected along the direction of the input probe beam as  a function of the mirror position. The signal is clearly modulated with nearly 100\% visibility and shows a $\sim4$x enhancement with respect to the same measurement performed without the mirror. Interestingly, the back-reflected signal is not a pure phase conjugated signal as one may have expected given the backward propagation direction (with respect to the probe). Indeed, we should also expect a negatively refracted beam however, comparing  the experimental layouts and beam arrangement shown in Figs.~\ref{mirror1}(a) and \ref{layout}(a), we should expect the negatively refracted beam to be reflected from the mirror and overlap, after passing through the graphene film, with the phase conjugated beam. We were able to verify that this is indeed the case by slightly detuning the angle of the mirror: in this way the negatively refracted and phase conjugated beams are emitted with slightly different angles and appear as two separate beams on a CCD camera placed at the photodetector position, as shown in Fig.~\ref{mirror1}(c) (the negatively refracted beam has a $\sim50\%$ lower intensity due to the absorption in the graphene film). Remarkably, as a result of the fact that they are generated in phase, these two signals sum coherently when they are overlapped, giving an enhanced signal as shown in CCD image inset in Fig.~\ref{mirror1}(a).\\

Finally, we show that this configuration may be simplified and used to observe a new kind of signal. Similarly to the case of single beam phase conjugation, where the two pumps and the seed  originate from the same beam \cite{Feinberg}, the retro-reflecting mirror is now placed at a slight angle of 3 deg and only a single pump beam (we block the probe beam) is used, see Fig.~\ref{setup3}(a). This configuration is a folded version of the second example shown here, i.e. generation of a negative refracted beam, using only two input beams incident at an angle albeit from opposite sides of the graphene film. We therefore expect here too a negatively refracted beam that will appear as reflected signal as shown schematically in Fig.~\ref{setup3}(b). Conservation of the transverse momentum dictates that the DFWM will appear reflected at an angle of $-2\theta$, where $\theta$ is the incident pump beam angle relative to the mirror normal. In other words, this beam appears as a ``negatively reflected'' beam, whose image, measured with a CCD camera, is shown in  Fig.~\ref{setup3}(c).\\

Two-dimensional materials exhibit unique features in the way they interact with light. Coherent enhancement or suppression of linear optical properties such as absorption \cite{zheludev}, scattering \cite{rao} and reflection \cite{zheludev_refl} are an example of the additional control enabled by the reduced dimensionality.  Here we have shown that the third order nonlinearity in a 2D material  exhibits an additional unique feature, namely it may be coherently controlled by modulating the phase of the input pump beam. This phase-dependence enables a series of optical wave mixing configurations that either optimise previously existing geometries or allow completely new possibilities. By fine tuning the relative phase of the pump beams, it is possible to coherently control and modulate the DFWM signals. The 2-dimensional nature of the material also implies that only one pump beam is required to generate a phase-conjugated beam and also a negatively refracted beam. Moreover, by choosing a geometry such that the beam diameters are of the same or order or smaller than the transverse interference pattern (controlled by the relative pump-probe angle), it is possible to observe full coherent control/modulation of the DFWM signals even with only two input beams. Finally, we showed how a reflective surface placed behind the 2-dimensional film at a small angle, acts as a nonlinear mirror that can generate a ``negatively reflected'' beam. \\
Beyond the fundamental implications of the phase-dependence of the nonlinearity in two-dimensional media, these ideas and others may find applications for example in the field of imaging. Perfect imaging has been demonstrated using two nonlinear films, e.g. metamaterial films, placed at a close distance and each individually pumped with two counterpropagating beams, i.e. with a total of 4 beams. The results here show that only one pump for each film is required and in principle, only one pump beam impinging on both films should be necessary to achieve the same perfect imaging results. Another possibility could be the coherent control of the nonlinearity to perform phase-contrast imaging where the phase of the probe{/pump} beam is directly mapped onto a corresponding intensity pattern in the phase-conjugated and/or negatively refracted beam. {Similarly, phase coherent control of the polarisation of the phase-conjugated and/or negative refracted beam may be achieved considering the inverse Faraday effect configuration, where the nonlinear medium is excited by a circularly polarised pump\cite{ife1,ife2}}.
\\

D.F. acknowledges financial support from the European Research Council under the European Union’s Seventh Framework Programme (FP/2007-2013)/ERC GA 306559 and EPSRC (UK, Grant EP/J00443X/1). M.C. acknowledges the support from the People Programme (Marie Curie Actions) of the European Union’s Seventh Framework Programme (FP7/2007-2013) under REA grant agreement n. 299522.

\end{document}